# ON THE LOCALITY PRINCIPLE KEEPING IN AHARONOV-BOHM EFFECT

Alexander Gritsunov and Natalie Masolova
*Dept. of Electronic Engineering*
Kharkiv National University of Radio Electronics, Ukraine
gritsunov@gmail.com; gritsunov@list.ru

*Abstract* — The locality principle fulfillment in the Aharonov-Bohm (AB) effect is analyzed from the point of view of a self-sufficient potential formalism based on so-called gradient hypothesis in electrodynamics. The "magnetic" kind of AB effect is examined (as the quantum charged particle moves to an infinitely long solenoid with a permanent current), and no locality principle violation recognized if the gradient hypothesis is used. A conclusion is made that AB effect is no longer a physical and electrodynamic "paradox".

## I. Introduction

Development of the nanotechnology is important and well-promising branch of contemporary engineering. However, some inveterate physical and electrodynamic problems like Aharonov-Bohm (AB) effect [1] interpretation may complicate the abovementioned process.

As numerous publications reveal, there is no way to reconcile AB effect with the locality principle using the Faraday-Maxwell electromagnetic (EM) field theory [2]. A field may be encapsulated in a superconductive case, but the wavefunction phase of a quantum particle moving around this case is still sensitive to the field value.

Self-sufficient potential formalism [3] proposes a new interpretation of the EM interactions. There is no EM field – all EM phenomena are natural and forced oscillations of the Minkowski space-time considered as a distributed EM oscillating system. The 4-vector of EM potential $\mathbf{A}(t,x,y,z)$ is a set of its generalized coordinates describing local deviation of the system from an "unperturbed" (electrically neutral) state. Any part of the EM oscillating system cannot be "insulated" with any conductor, EM potential relaxes only with a distance.

Let's consider how the AB effect problem is solved from the position of the potential formalism.

## II. Vortex and Gradient Hypotheses

The Lagrange function 3-density $\lambda(t,x,y,z)$ for a system of charged particles coupled with EM interactions consists of three terms: $\lambda = \lambda^P + \lambda^I + \lambda^S$. $\lambda^P$ is a "mechanical" term for the particles [2]. $\lambda^I = \mathbf{A} \cdot \mathbf{j}$ is an "interaction" term between the particles and EM field (in field formalism) or the distributed EM oscillating system (in potential formalism). $\lambda^S$ describes the EM field or the oscillating system itself respectively and may be postulated using two different hypotheses: vortex

$$\lambda^S = -\frac{1}{2\mu_0}(\nabla \times \mathbf{A})^2$$

that is principal for the field formalism [2]; and gradient

$$\lambda^S = -\frac{1}{2\mu_0}\left[(\nabla A_t)^2 - (\nabla A_x)^2 - (\nabla A_y)^2 - (\nabla A_z)^2\right]$$

that is a base for the potential formalism [3]. In the formulas above, $\mathbf{j}(t,x,y,z)$ is the current density 4-vector; $\nabla \times \mathbf{A}$ is tensor of 4-curl of $\mathbf{A}$ [4]; $\nabla A_r$ is the 4-gradient of $A_r$; $r$ is a generic symbol for $t \equiv ct$, $x$, $y$, or $z$.

The locality principle expresses itself in the energy-momentum flow continuity law. This flow is described by the energy-momentum density tensor $[w]$ also consisting of three terms: $[w] = [w^P] + [w^I] + [w^S]$.

"Mechanical" tensor for particles $[w^P]$ is considered in [2]. Components of "interaction" between the particles and the distributed EM oscillating system tensor are as $w^I_{rr'} = A_r j_{r'}$. Components for the oscillating system itself tensor are calculated from $\lambda^S$ as [2]:

$$w^S_{rr'} = g_{rr}\sum_{r''}\frac{\partial A_{r''}}{\partial r}\frac{\partial \lambda^S}{\partial(\partial A_{r''}/\partial r')} - g_{rr'}\lambda^S,$$

where $[g]$ is the diagonal metric tensor for the pseudo Euclidian space. As known, $[w^S]$ is originally asymmetric for the vortex hypothesis. Therefore, special symmetrization procedure must be applied to guarantee the angular momentum conservation. Such procedure is described in [2] for a specific case $\mathbf{j} \equiv 0$, but considered in [4] general case of $\mathbf{j} \neq 0$ is of particular interest. It is shown that $[w^I]$ vanishes from $[w]$ after the symmetrization. The symmetrization displaces EM energy and momentum within a closed system by "taking away" from particles and "spreading" over the space where $\nabla \times \mathbf{A}$ components are non-zero. Such "non-physical" relocation of energy-momentum is the prime cause of some EM "paradoxes" [5] and also results in a seeming violation of locality principle in AB effect.

On the contrary, $[w^S]$ is symmetric originally for the gradient hypothesis, no symmetrization needs:

$$w^S_{rr'} = -\frac{g_{rr}g_{r'r'}}{\mu_0}\left(\frac{\partial A_t}{\partial r}\frac{\partial A_t}{\partial r'} - \frac{\partial A_x}{\partial r}\frac{\partial A_x}{\partial r'} - \frac{\partial A_y}{\partial r}\frac{\partial A_y}{\partial r'} - \frac{\partial A_z}{\partial r}\frac{\partial A_z}{\partial r'}\right)$$
$$+ \frac{g_{rr'}}{2\mu_0}\left[(\nabla A_t)^2 - (\nabla A_x)^2 - (\nabla A_y)^2 - (\nabla A_z)^2\right], \qquad (1)$$

so $[w^I]$ remains in $[w]$. Generally, this ensures the locality principle keeping for any kind of AB effect ("electric" or "magnetic"), as just $[w^I]$, not $[w^S]$ components appear in the expression for the quantum particle wavefunction phase incursion along a closed 4-loop $L$ [5]:

$$\Delta\varphi = \frac{q}{\hbar}\oint_L \mathbf{A}(t,x,y,z) \cdot d\mathbf{l}.$$

The confirmation of this thesis is given in [6] for the "electric" AB effect. Let's demonstrate the locality principle conservation for the "magnetic" kind of AB effect.

## III. Solenoid and Charged Particle

Let's consider an infinitely long along $z$ solenoid with radius $r_0$ and linear density of permanent azimuthal current $i_0$ (see Fig. 1). Solutions for the solenoid potential 4-

vector $\mathbf{A}_0(x,y)$ components from the outside are

$$A_{0x} = -\frac{1}{2}\mu_0 i_0 r_0^2 \frac{y}{x^2+y^2}; \quad A_{0y} = \frac{1}{2}\mu_0 i_0 r_0^2 \frac{x}{x^2+y^2}. \quad (2)$$

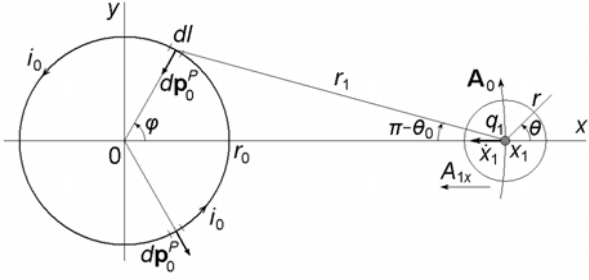

*Fig. 1. A charged hair moves to a solenoid*

An infinitely long along $z$ charged hair with linear density of charge $q_1$ approaches to the solenoid along $x$ with the velocity $\dot{x}_1 = dx_1/dt = \text{const} < 0$ (2D system is considered for simplicity). The linear densities of "interaction" momentum $\mathbf{P}_1^I(t)$ components for the hair are

$$P_{1x}^I = q_1 A_{0x}[x_1(t),0] = 0; \quad P_{1y}^I = q_1 A_{0y}[x_1(t),0] = \frac{\mu_0 i_0 q_1 r_0^2}{2 x_1(t)},$$

so their derivatives with respect to $t$ are

$$\frac{dP_{1x}^I}{dt} = 0; \quad \frac{dP_{1y}^I}{dt} = -\frac{\mu_0 i_0 q_1 r_0^2 \dot{x}_1}{2 x_1^2}. \quad (3)$$

Let's surround the hair with imaginary cylindrical surface of negligibly small radius $r$ (see Fig. 1). The "mechanical" momentum of hair $\mathbf{P}_1^P$ as well as the momentum of the EM oscillating system within the surface $\mathbf{P}_1^S$ linear densities both are unvaried. Therefore, the momentum flow continuity law can be written as

$$-\frac{dP_{1x}^I}{dt} = \frac{1}{c}\int_0^{2\pi} w_{xr}^S r d\theta; \quad -\frac{dP_{1y}^I}{dt} = \frac{1}{c}\int_0^{2\pi} w_{yr}^S r d\theta, \quad (4)$$

where

$$w_{xr}^S(r,\theta) = w_{xx}^S(r,\theta)\cos\theta + w_{xy}^S(r,\theta)\sin\theta;$$
$$w_{yr}^S(r,\theta) = w_{yx}^S(r,\theta)\cos\theta + w_{yy}^S(r,\theta)\sin\theta$$

are surface densities of flows of the EM oscillating system momentum $x$ and $y$ components through the imaginary cylindrical surface in the outward normal direction.

Non-zero components of the moving hair potential $\mathbf{A}_1(r)$ at the surface are

$$A_{1t} = -\frac{\mu_0 c q_1}{2\pi}\ln r + \text{const}; \quad A_{1x} = -\frac{\mu_0 c q_1}{2\pi}\dot{x}_1 \ln r + \text{const}.$$

Using (1) in assumption $|\dot{x}_1| \ll 1$ and integrating over the imaginary cylindrical surface perimeter, we obtain:

$$\frac{1}{c}\int_0^{2\pi} w_{xr}^S r d\theta = 0; \quad \frac{1}{c}\int_0^{2\pi} w_{yr}^S r d\theta = \frac{\mu_0 i_0 q_1 r_0^2 \dot{x}_1}{2 x_1^2}. \quad (5)$$

Comparing (3) and (5) one can see that (4) is fulfilled. While a charged particle approaches to the solenoid, $w_{\xi\xi'}^S$ are non-zero from the outside ($\xi$ is a generic symbol for $x$, $y$, or $z$), so EM momentum flows to the particle through the distributed EM oscillating system. The locality principle is kept for the gradient hypothesis.

Let's examine whether the total momentum and angular momentum conservation laws are kept for the system. Only $dP_{1y}^I/dt \neq 0$ for the hair (3), so this increment in the hair momentum must be compensated by equal decrement in the solenoid momentum (as the EM oscillating system momentum is invariable). The "magnetic field" produced by the moving hair (i.e., $\nabla \times \mathbf{A}_1$) interacts with the solenoid current $i_0$ causing permanent increase $d\mathbf{p}_0^P$ of the solenoid "mechanical" momentum linear density per its wall fragment $dl$ (see Fig. 1). The derivative with respect to $t$ of radial component of $d\mathbf{p}_0^P$ is

$$\frac{dp_{0r}^P}{dt} = \frac{\mu_0 i_0 q_1 r_0 \dot{x}_1}{2\pi r_1(\varphi)}\sin\theta_0(\varphi)d\varphi.$$

After integrating over all perimeter of the solenoid, expressions for the rate of the solenoid total "mechanical" momentum linear density $\mathbf{P}_0^P(t)$ growth are obtained:

$$\frac{dP_{0x}^P}{dt} = 0; \quad \frac{dP_{0y}^P}{dt} = \frac{\mu_0 i_0 q_1 r_0^2 \dot{x}_1}{2 x_1^2}. \quad (6)$$

A conclusion can be made comparing (6) with (3) that $d(\mathbf{P}_0^P + \mathbf{P}_1^I)/dt \equiv 0$, i.e. the total momentum conservation law for the system is fulfilled.

Because all $d\mathbf{p}_0^P$ are oriented strictly towards or outward the solenoid axis, total "mechanical" angular momentum of the solenoid about $z$ axis is unchanged. Therefore, the "interaction" angular momentum linear density of the hair about the same axis $q_1 x_1 A_{0y}(x_1, 0)$ must also be constant. As it can be seen from (2), this condition is also satisfied. There are no violation of the conservation laws in the examined system.

## IV. Conclusion

The AB effect has simple and "elegant" interpretation in the self-sufficient potential formalism using the gradient hypothesis. No conflict with the locality principle appears. As a result, an assumption can be made that the inveterate problem of AB effect as some physical and electrodynamic "paradox" no longer exists.

## V. References

[1] *Aharonov, Y. and D. Bohm*, "Significance of electromagnetic potentials in quantum theory," *Physical Review*, vol. 115, no. 3, pp. 485-491, 1959.
[2] *Landau, L. D. and E. M. Lifshitz*, Course of Theoretical Physics, *The Classical Theory of Fields*, vol. 2, Butterworth-Heinemann, Oxford, 1975.
[3] *Gritsunov, A. V.*, "Self-sufficient potential formalism in describing electromagnetic interactions," *Radioelectronics and Comm. Systems*, vol. 52, no. 12, pp. 649-659, 2009.
[4] *Gritsunov, A. V. and N. V. Masolova,* "On the adequacy of vortex hypothesis in the self-sufficient potential formalism," *Mag. of Sci. Trans. of Nakhimov Naval Acad.*, no. 4(12), 225-235, 2012 (in Russian).
[5] *Feynman, R. P., R. B. Leighton and M. Sands,* The Feynman Lectures on Physics, *Mainly Electromagnetism and Matter*, vol. 2, Addison-Wesley, Reading, MA, 1964.
[6] *Gritsunov, A. V.,* "Why Aharonov-Bohm effect does not violate locality principle," *in 14th Int. Vacuum Electronics Conf. (IVEC 2013)*, Paris, France, 2013 (to be published).